\documentclass[aps,prd,preprint,groupedaddress,preprintnumbers,floatfix]{revtex4}
\usepackage{epsf}
\usepackage{epsfig}
\usepackage{graphics}
\usepackage{subfigure}
\usepackage{xspace}
\newcommand{\ra} {\rightarrow}

\newcommand{\be}{\begin{equation}}
\newcommand{\ee}{\end{equation}}

\newcommand{\bea}{\begin{eqnarray}}
\newcommand{\eea}{\end{eqnarray}}
\newcommand{\beanon}{\begin{eqnarray*}}
\newcommand{\eeanon}{\end{eqnarray*}}
\newcommand{\ba}{\begin{array}}
\newcommand{\ea}{\end{array}}
\newcommand{\bd}{\begin{description}}
\newcommand{\ed}{\end{description}}
\newcommand{\bi}{\begin{itemize}}
\newcommand{\ei}{\end{itemize}}
\newcommand{\ben}{\begin{enumerate}}
\newcommand{\een}{\end{enumerate}}
\newcommand{\bc}{\begin{center}}
\newcommand{\ec}{\end{center}}

\newcommand{\VV}{\mbox{${\mathrm V}{\mathrm V}$}\xspace}

\newcommand{\VL}{\mbox{${\mathrm V_L}$}\xspace}

\newcommand{\pt}{\mbox{${\mathrm p_T}$}\xspace}

\newcommand{\eqns}[2]{Eqs.(\ref{#1}--\ref{#2})}

\newcommand{\tbn}[1]{Tab.~\ref{#1}}

\newcommand{\fig}[1]{Fig.~\ref{#1}}

\newcommand{\sects}[2]{Sects.~\ref{#1},~\ref{#2}}

\def\pl #1 #2 #3 {{\it Phys.~Lett.} {\bf#1} (#2) #3}   
\def\np #1 #2 #3 {{\it Nucl.~Phys.} {\bf#1} (#2) #3}
\def\zp #1 #2 #3 {{\it Z.~Phys.} {\bf#1} (#2) #3}
\def\pr #1 #2 #3 {{\it Phys.~Rev.} {\bf#1} (#2) #3}
\def\prep #1 #2 #3 {{\it Phys.~Rep.} {\bf#1} (#2) #3}
\def\prl #1 #2 #3 {{\it Phys.~Rev.~Lett.} {\bf#1} (#2) #3}
\def\intj #1 #2 #3 {{\it Int. J. Mod. Phys.} {\bf#1} (#2) #3}
\def\mpl #1 #2 #3 {{\it Mod.~Phys.~Lett.} {\bf#1} (#2) #3}
\def\rmp #1 #2 #3 {{\it Rev. Mod. Phys.} {\bf#1} (#2) #3}
\def\cpc #1 #2 #3 {{\it Comp. Phys. Commun.} {\bf#1} (#2) #3}
\def\epj #1 #2 #3 {{\it Eur. Phys. J.} {\bf#1} (#2) #3}
\def\jhep #1 #2 #3 {{\it JHEP} {\bf#1} (#2) #3}

\begin{document}

\preprint{DFTT 12/2006}

\title{Isolating Vector Boson Scattering at the LHC:
            gauge cancellations and the
            Equivalent Vector Boson Approximation
            vs complete calculations.}

\author{Elena Accomando}
\email{accomand@to.infn.it}

\author{Alessandro Ballestrero}
\email{ballestr@to.infn.it}
\author{Aissa Belhouari}
\email{belhouar@to.infn.it}
\author{Ezio Maina}
\email{maina@to.infn.it}

\affiliation{
          INFN, Sezione di Torino and
  Dipartimento di Fisica Teorica, Universit\`a di Torino\\
  Via Giuria 1, 10125 Torino, Italy}

\thanks{
E.A. is supported by the Italian Ministero dell'Istruzione, 
dell'Universit\`a e della Ricerca (MIUR) under contract Decreto MIUR 
26-01-2001 N.13 ``Incentivazione alla mobilit\`a di studiosi stranieri ed 
italiani residenti all'estero''.\\
Work supported by MIUR under contract 2004021808\_009.
}

\begin{abstract}
We have studied the possibility of 
extracting the $W^+W^-\rightarrow W^+W^-$ signal using the process 
$us\rightarrow cdW^+W^-$ as a test case.
We have investigated numerically the strong gauge 
cancellations between signal and irreducible background,
analysing critically the reliability of the Equivalent Vector Boson
Approximation which is commonly used to define the 
signal.
Complete matrix elements are necessary to study Electro--Weak Symmetry Breaking
effects at high $WW$ invariant mass.
\end{abstract}

\maketitle

%\keywords{abc}

\section{Introduction}
\label{sec:intro}
The mechanism of Electro--Weak Symmetry Breaking (EWSB) will be 
one of the primary topics to be investigated at the
LHC,
both through direct searches for the Higgs boson and careful analysis of
boson boson scattering processes.
Detailed reviews and extensive bibliographies can be found in 
Refs.\cite{atl:99atl,Assamagan:2004mu,Djouadi:2005gi}.
The nature of the interaction between longitudinally polarized vector bosons
and the Higgs mass, or possibly the absence of
the Higgs particle, are strongly related: if a relatively light Higgs exists
then
the \VL's  are weakly coupled, while they are strongly interacting if the Higgs
mass is large or the Higgs is nonexistent \cite{Chanowitz:1998wi}.

It should be noted that the Goldstone theorem and the Higgs mechanism do not
require
the existence of elementary scalars. It is conceiveable and widely discussed in
the literature that bound states are responsible for EWSB.

At the LHC no beam of on shell EW vector
bosons will be available. Incoming quarks will
emit spacelike virtual bosons which will then scatter among themselves and 
finally decay.
These processes have been scrutinized since a long time in order to uncover the
details of the EWSB mechanism in this realistic setting
\cite{Duncan:1985vj,Dicus:1986jg,Gunion:1986gm}.
Naively one expects that at large boson boson invariant masses the diagrams
containing vector boson fusion subdiagrams should dominate the total
cross section
while the offshellness of the bosons initiating the scattering process should
become less and less relevant. 
Together with the keen interest for Higgs production in vector boson fusion,
this has lead to the development of the Equivalent Vector Boson
Approximation (EVBA) \cite{Dawson:1984gx,Kane:1984bb, Lindfors:1985yp}.
The EVBA provides a particularly simple and
appealing framework in which the cross section for the
full process is approximated by the convolution
of the cross section for the scattering of on shell vector bosons times
appropriate
distribution functions which can be interpreted as the probability of the
initial state quarks to emit the EW bosons which then interact.

This approach relies on the neglect of all diagrams which do not include
boson boson scattering subdiagrams and on a suitable on-shell projection
for the scattering set of diagrams.
Since the approximate boson boson interaction is expressed in terms
of on shell particles
it is straitforward to separate the different boson polarizations.
It is well known that the set of scattering diagrams is not separately gauge
invariant while both the on shell amplitude and the distribution functions
which appear in the EVBA are gauge independent.

In \cite{Kleiss:1986xp} it has been shown that when
vector bosons are allowed to be off mass shell in boson boson scattering,
the amplitude grows faster with energy compared with the amplitude for
on shell vectors. 
Subsequently, it has been pointed out in \cite{Kunszt:1987tk} that the
problem of bad high energy behaviour of $WW$ scattering diagrams
can be avoided by the use of the Axial gauge. 
Recently $WZ$ production at hadron collider \cite{Accomando:2006iz} has been
analyzed in Axial gauge with very encouraging results.
It should be noted that the results in \cite{Kunszt:1987tk} have been obtained
under the assumption that the transverse momenta of the produced $W$'s are
of the order of the Higgs mass and that each are much larger than the $W$ mass.
It is therefore not obvious to what degree the conclusions of
Ref.\cite{Kunszt:1987tk} can be applied in the LHC environment, particularly
for light Higgs masses as preferred by global SM fits.

The EVBA has some undesirable features: a number of
unphysical cuts need to be introduced to tame the singularities generated by the
onshell projection, which are absent from the exact amplitude.
In the literature, a number
of comparisons of exact and EVBA calculations have appeared with conflicting
results \cite{Cahn:1984tx,Willenbrock:1986cr,Gunion:1986gm}

For these reasons in the present paper we have critically examined the
role of gauge
invariance in \VV -fusion processes and the reliability of the EVBA in
describing them. We would like to determine
regions in phase space, at least in a suitable gauge,
which are dominated by the scattering set of diagrams.

If this was the case then it would be possible employ this set
of diagrams to define a ``signal'',
%as was done with {\tt CC03} at LEP 2, 
that is a pseudovariable which could be used to compare the results from the
different collaborations. The signal is not necessarily directly observable
but it should be possible to relate it via Monte Carlo to measurable quantities.
If such a definition is to be useful it must correspond as closely as possible
to the process which needs to be studied and the Monte Carlo corrections
must be small.
If instead, as we are led to believe by the results shown in the following,
the VV scattering diagrams
do not constitute the dominant contribution in any gauge or phase space region,
there is no substitute to the complete amplitude for studying boson fusion
processes at the LHC.
We cannot claim to have examined all possible gauge scheme,
however we have studied the
most commonly used $\xi$--type gauges and the implementation of the Axial gauge
proposed in \cite{Kunszt:1987tk}.
The results obtained with the full amplitude show that the mechanism of EWSB can
indeed be investigated even without separating out the scattering set of
diagrams.

\section{VV -Scattering and gauge invariance}
\label{sec:GaugeInv}
\begin{figure}[htb]
\begin{center}
%\mbox{\epsfig{file=uscdww.pdf,width=12.cm}}
\mbox{\epsfig{file=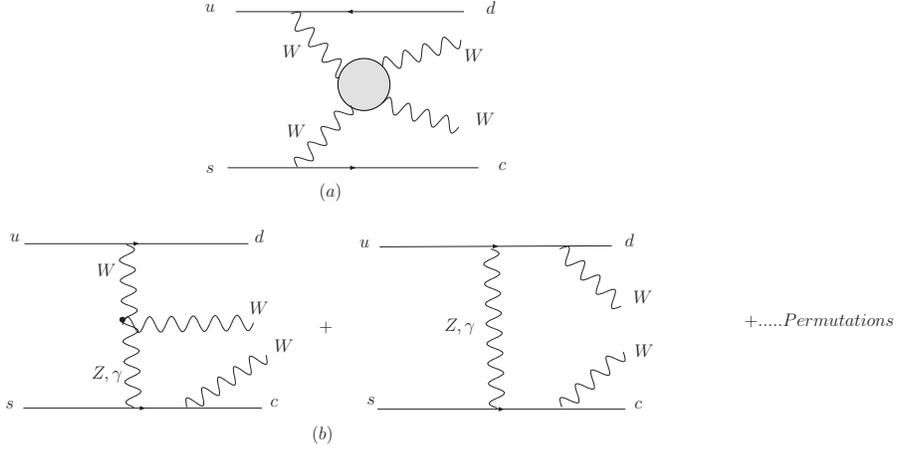,width=12.cm}}
\caption{\footnotesize Main diagram topologies for
the process $us \rightarrow cd W^+W^-$}
\label{uscdww}
\end{center}
\end{figure}

In order to study the implications of gauge invariance in \VV -fusion processes
we have concentrated on the specific process $us\ra cd W^+W^-$ in proton
proton collisions at the LHC. The corresponding Feynman diagrams can be
classified in two different sets (\fig{uscdww}),
the boson boson fusion one
(\fig{uscdww}a) and all
the rest (\fig{uscdww}b) in which at least one final W is emitted
by a fermion line. The two sets are not separately gauge invariant.
We have started our analysis from the contribution  of the boson boson
fusion diagrams and their interference with the remaining ones.
We have considered the Unitary, Feynman and Landau gauge. Among these, we find
that the Feynman
gauge is the one which minimizes the cancellations between set (a) of
\fig{uscdww} and the non scattering diagrams.
We have also performed the calculation in the  Axial gauge $n_{\mu}A^{\mu}=0$
within the scheme proposed in Ref.\cite{Kunszt:1987tk}.
In Appendix.A we have collected the main corresponding Feynman rules.   
We have tried
different $n_\mu$ gauge vectors and we find that the choice
$n_\mu = (1 , 0 , 0 , 1)$
(The incoming protons propagate along the z axis) is the best one.
For brevity we will
refer to this framework as Axial gauge in the following.
We present results for the Unitary, Feynman and Axial gauge, showing the
contribution of all diagrams ($all$), of the $WW$ scattering diagrams ($WW$) 
together with
their ratio $WW/all$. When this ratio is significantly greater than one
the contribution of non scattering diagrams is of the same order
as the contribution of the $WW$ scattering ones and important cancellations
take place between the two sets.

In \tbn{crosx_noh} the total cross sections
and their ratios are computed in the limit of infinite Higgs mass. This limit
will also be referred to in the following as noHiggs case.
We find that the Axial gauge 
is the one in which the interferences are least severe
with a ratio of about 2, while the ratios for the Unitary and Feynman gauge
are 358 and 13 respectively. The inclusion of a light
Higgs ($M_h$=200 GeV) does not improve matters:
on the contrary the ratios become even larger as shown in \tbn{crosx_h}
for $M_h$=200 GeV, doubling for the Unitary and Feynman gauge

\begin{table}[htb]
\begin{center}
\begin{tabular}{|l|c|c|c|c|} 
\hline
%& \multicolumn{1}{|c|}{No-Higgs} \\
   &   $\sigma(pb)$ All diagrams  & $\sigma(pb)$ $WW$ diagrams & ratio $WW/all$ \\
\hline
Unitary gauge & 1.86 10$^{-2}$    &  6.67 & 358\\
\hline
Feynman gauge & 1.86 10$^{-2}$      & 0.245 & 13\\
\hline
Axial gauge  &  1.86 10$^{-2}$    & 3.71 10$^{-2}$ & 2 \\
\hline
\end{tabular}
\caption{\footnotesize  $WW$ diagrams and complete set of diagrams cross
sections and their ratios computed in different gauges without Higgs
contribution. We have used the CTEQ5 Pdf set with scale $M_W$}
\label{crosx_noh}
\end{center}
\end{table}
%\vspace{-0.5cm}
\begin{table}[htb]
\begin{center}
\begin{tabular}{|l|c|c|c|c|} 
\hline
%& \multicolumn{1}{|c|}{No-Higgs} \\
   &   $\sigma(pb)$ All diagrams  & $\sigma(pb)$ $WW$ diagrams & ratio $WW/all$ \\
\hline
Unitary gauge & 8.50 10$^{-3}$   &  6.5 & 765\\
\hline
Feynman gauge & 8.50 10$^{-3}$ & 0.221 & 26 \\
\hline
Axial gauge  & 8.50 10$^{-3}$ & 2.0 10$^{-2}$  & 2.3 \\
\hline
\end{tabular}
\caption{\footnotesize  $WW$ diagrams and complete set of diagrams cross sections
and their ratios computed in different gauges with a $M_h$ =200 GeV Higgs 
and $M(WW) >$ 300 GeV .}
\label{crosx_h}
\end{center}
\end{table}

\begin{figure}[tbh]
\begin{center}
%\mbox{\epsfig{file=Total_WW_newv2.pdf,width=8cm}
%\epsfig{file=ratio_v2.pdf,width=8cm}}
\mbox{\epsfig{file=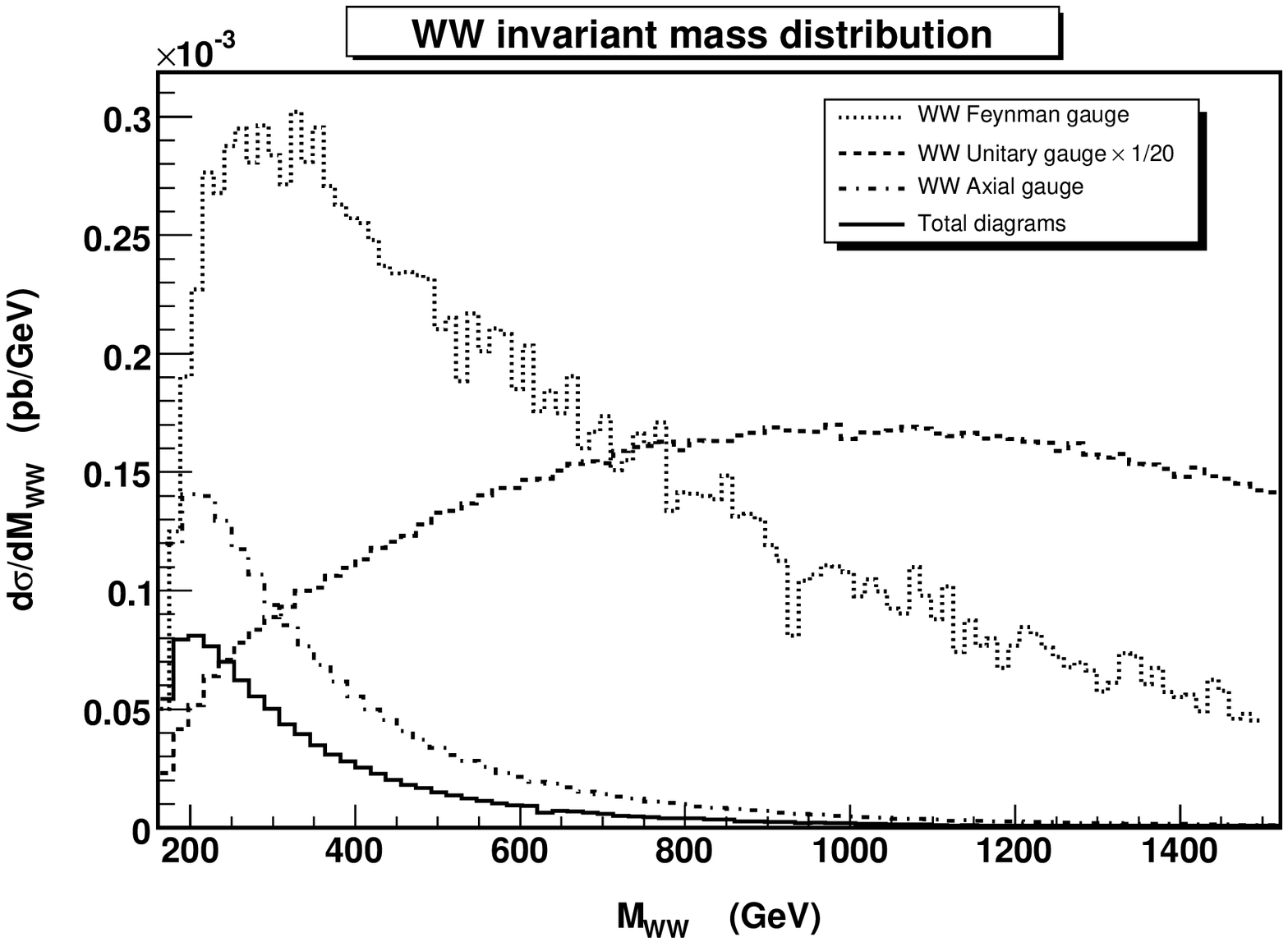,width=8cm}
\epsfig{file=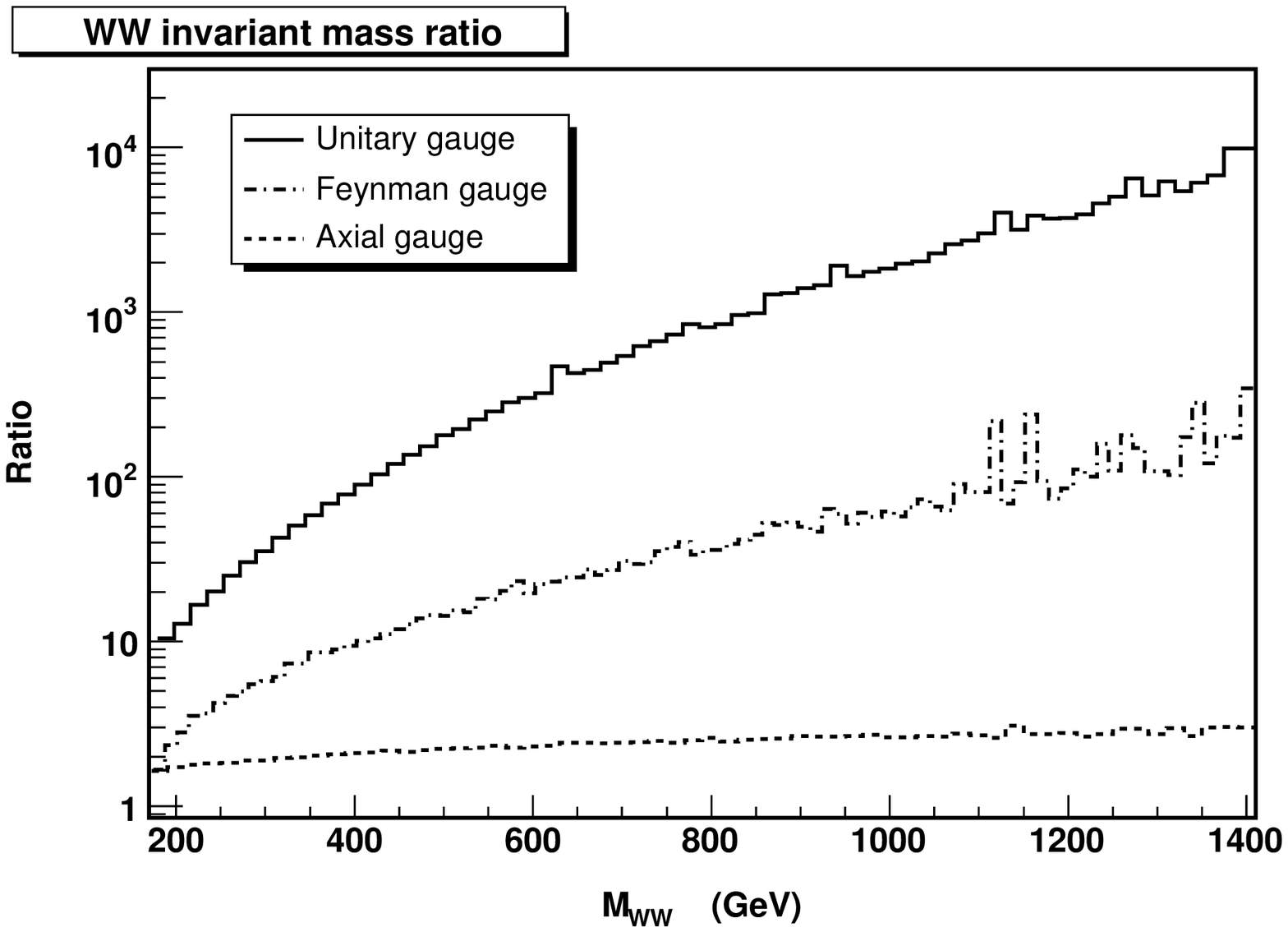,width=8cm}}
\caption{ Distribution of $d\sigma/dM_{WW}$ for the process
$PP\rightarrow us \rightarrow cdW^+W^-$
for All diagrams, $WW$ diagrams and their ratio in Unitary, Feynman and Axial
gauge in the infinite Higgs mass limit.
The Unitary gauge data in the left hand plot
have been divided by 20 for better presentation.}
\label{wwMassDistGauge}
\end{center}
\end{figure}

The comparison of total cross sections is only a preliminary step. It provides
the general behavior but it cannot give informations on the different regions
of phase space. For this reason we have evaluated the
distribution of different kinematical variables with the goal 
of finding regions where the interferences are not important. If such regions
exist it will be possible to define a set of cuts which allow the extraction of
the $WW$ scattering amplitude. \fig{wwMassDistGauge} shows the distributions
of the diboson invariant mass
for the complete set of diagrams and for the $WW$ diagrams only
together with the ratio $(d\sigma(WW)/dM)/(d\sigma(All)/dM)$ . 
We have chosen the vector boson pair invariant mass distribution
as a prototype but the same conclusions are reached with all
other variables. The results of Fig. \ref{wwMassDistGauge}
have been obtained for a very large Higgs mass but the
general behaviour is not modified by the inclusion of a light Higgs.
The distribution obtained with the full set of diagrams
and with the $WW$ fusion set only are quite different in Unitary and
Feynman gauge. The ratio is also large over the whole interval and
especially in the region of high invariant mass which is the most important one
for
EWSB studies. Again the Axial gauge gives the best result, with distributions
which have the same general shape in the two cases. The ratio in this
gauge remains however greater than 2, apart from
a very small region at low invariant mass. 

\begin{figure}[ht]
\begin{center}
\mbox{\epsfig{file=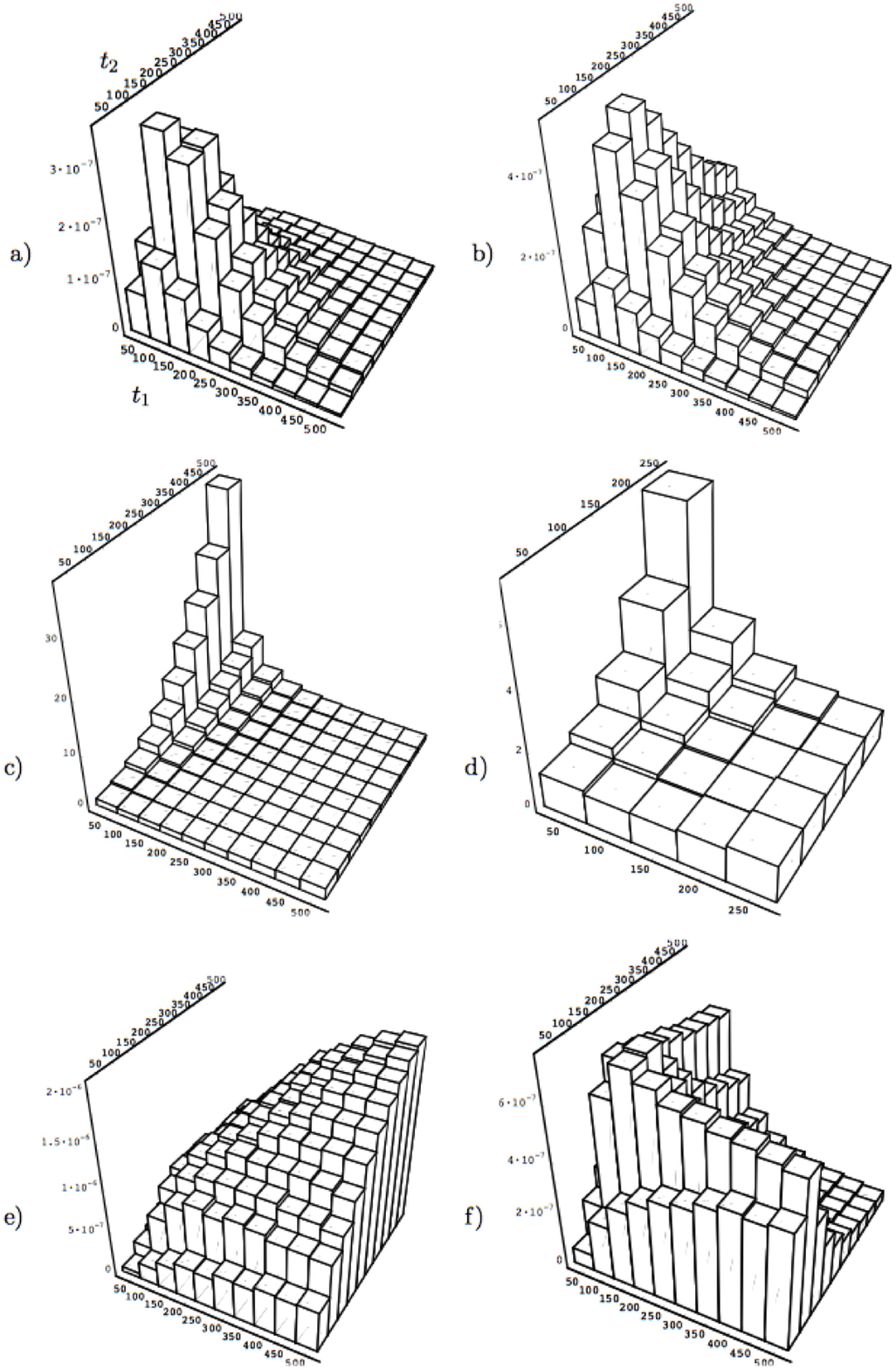,width=14cm}}
\caption[]{ Double differential distributions of $d\sigma/dt_1dt_2$ 
$(pb/GeV^2) $  with $t_{1,2}=\sqrt{-(p_{u,s}-p_{d,c})^2}$ for the process 
$PP\rightarrow us \rightarrow dc W^+W^-$ , using All diagrams 
(a) or only the $WW$ fusion subset in Axial gauge (b).
The two plots in the central row represent the ratio $WW/all$ in the Axial gauge
as a function of $t_1,t_2$ 
for the total region (c) and for small $t$'s (d).
For comparison we also show the distributions of the $WW$ fusion subset 
in the Unitary (e) and Feynman (f) gauge. All invariants are in GeV.
}
\label{t1_t2_axi}
\end{center}
\end{figure}

We have completed our study by analyzing bi-dimensional distributions of
several pairs of kinematical variables. These double distributions allow us
to analyze in particular
the situation where the two incoming bosons have small virtualities
($t_{1,2}\ra 0$). In this region the $WW$ scattering diagrams are expected
to dominate.
The same behaviour found previously is observed: the effect of the interferences
is much more relevant in Unitary gauge with respect to the others. In general
the $WW$ fusion subset has a different behavior compared with the complete
calculation. 
In \fig{t1_t2_axi} we report
$d\sigma/dt_{1}dt_{2}$ in the Axial gauge where 
$t_{1,2} = \sqrt{-(p_{u,s}-p_{d,c})^2}$  are the square root of the absolute
value of the invariant masses of the incoming 
off shell W's. 
The corresponding ratio distributions (plots (c) (d)) show that even in the
limited region $0 <t_{1,2}<250$ GeV, the $WW$ contribution alone does not
provide a realistic description of the complete set of diagrams.
Even in Axial gauge significant cancellations take
place. The ratio is highly asymmetric in the $t_{1}-t_{2}$ plane, reflecting its
sensitivity to the choice of the Axial gauge axis (Recall that in this case the
gauge vector is along the $z$--axis)
and $WW$ scattering diagrams alone provide only a very rough estimate of
the cross section. It is however possible to find regions, typically
for $t_{1,2} > 200$ GeV where the cross section is greatly reduced,
where the ratio of the two results is around one.
For comparison we also show the distributions obtained from the $WW$ fusion
subset in the Unitary (e) and Feynman (f) gauge whose shape, not to mention the
normalization, is completely different from the results obtained from the full
amplitude.

From this sample of results one can conclude that the $WW$ scattering diagrams
do not constitute the dominant contribution in any phase space region
for the gauges we have examined.
For the Axial gauge, which in various cases shows ratios of order 2,
we remark that these have been obtained only with a particular 
time like gauge vector $n_\mu = (1, 0, 0, 1)$. For any other space-like 
$n_{\mu}$, the interferences are
so important that even the numerical integration becomes difficult for the $WW$
diagrams set whereas no problem occurs for the full set of diagrams.
So we did not succeed in finding a gauge vector of the type
suggested in \cite{Kunszt:1987tk} for which the cancellations are negligible.
Even with our best choice of $n_\mu$ the distributions of the different
kinematical variables show the presence of large interferences between the two
subset of diagrams over most of  phase space, indicating that the $WW$ diagrams
are not dominating. As a consequence a question mark is put on the possibility
to isolate the $WW$ scattering contribution by restricting the calculations
to the corresponding diagrams. The reliability of approximation methods
based on such an approach becomes then suspicious. For this reason we have
completed our analysis by studying the Effective Vector Boson Approximation
applied to our prototype process.

\newpage
%\enlargethispage{2\baselineskip}

\clearpage

\section{The Effective Vector Boson Approximation}
\label{sec:EVBA}
The Effective Photon Approximation known
also as Weizs\"{a}cker-Williams approximation
\cite{vonWeizsacker:1934sx,Williams:1934ad}
has proved to be a useful tool in the study of photon-photon processes
at $e^+ e^-$ colliders.
Encouraged by this success the approach has been extended
to processes involving massive vector bosons
\cite{Dawson:1984gx, Kane:1984bb, Lindfors:1985yp} under the name of
Effective Vector Boson  Approximation.
The EVBA  has been first applied at hadron colliders 
in connection with Higgs production $pp\rightarrow H + X$  \cite{Cahn:1983ip}
and subsequently to vector boson processes of the type
$pp\rightarrow ( H\rightarrow V_3V_4 ) + X\rightarrow V_3 V_4  + X$
\cite{Gunion:1986gm,Chanowitz:1984ne}
in order to obtain EVBA
predictions for the production of a vector boson pair not necessarily
near the Higgs resonance and to study the strongly interacting
scenario.

In analogy to
the QED case, the application of EVBA to these processes consists in:

\ben
\item Restricting the computation to the vector boson scattering diagrams
neglecting diagrams of bremsstrahlung type.
\item  Projecting on-shell the momenta of the vector bosons which take part
to the scattering:
$q^2_{1,2} = M^2_{V_{1,2}}$. Here it is important to notice that contrary
to the $\gamma\gamma$ processes where the photon momentum can reach
the on shell
value $q^2_{1,2}= 0$, for the massive vector bosons the onshell point
$q^2_{1,2} = M^2_{V_{1,2}}$ is outside the accessible phase space
region $q^2_{1,2}\leq 0$.
\item Approximating the total cross section of the process
$f_1f_2\rightarrow f_3f_4V_3V_4$
as the convolution of the vector boson luminosities
$\mathcal {L}^{V_{1}V_{2}}_{Pol_{1}Pol_{2}}(x)$ with the on shell vector
boson scattering cross section: 
\een
\be\label{evba_cros}
\sigma(f_1f_2\rightarrow f_3f_4V_3V_4) =
\int\sum_{V_1,V_2}\sum_{Pol_{1}Pol_{2}}
{\mathcal{ L}^{V_{1}V_{2}}_{Pol_{1}Pol_{2}}(x)}
\sigma^{on}_{pol}(V_1V_2\rightarrow V_3V_4 ,xs_{qq})dx
\ee

Here $x=M(V_{1}V_{2})^2/s_{qq}$, while $M(V_{1}V_{2})$ is the vector boson pair
invariant mass and $s_{qq}$ is the partonic center of mass energy.

It is clear that this approximation provides a simplification from the
computational point of view and can exploit the properties of on shell
boson boson scattering.
In first applications of EVBA to vector boson scattering further
approximations have been adopted. The zero angle scattering approximation
has been used for the process $V_1V_2\rightarrow V_3V_4$ so the transverse momentum
of the incoming bosons has been neglected. The zero mass limit for the
vector boson mass $M_{V}\rightarrow 0$  has also been considered in the
computation of the luminosity.
In this early application of EVBA only the contribution from
longitudinal modes was considered while the contribution from transverse
states was neglected. The EVBA results depended strongly on the details of
the approximations made. The EVBA generally
overestimated the complete perturbative calculation, in some cases
by a factor 3 \cite{Godbole:1987iy}. Progressively, more refined and rigorous
formulations of EVBA have been proposed, avoiding as much as
possible the mentioned approximations, until a formulation
where no kinematical approximations are taken and all vector boson
polarization states as well as their interferences are taken into account
\cite{Kuss:1995yv}.
Inspired by the strategy of \cite{Kuss:1995yv}, we
have further improved the EVBA. We have used an
approach where not only all kinematic approximations are avoided but also
the luminosity computation is not needed. This allows in particular to keep
all final particle properties (momenta, angles...) as they would be in an
exact calculation contrary to the traditional approach where a
pre-integration is performed to obtain the vector boson luminosities.

\section{EVBA application to the  
$PP\rightarrow us \rightarrow dc W^+W^-$ process}
\label{sec:PP2usWW}
In \cite{Kuss:1995yv} a precise formulation of EVBA has been developed.
It is based on a factorization technique for analyzing Feynman diagrams
which leads to exact probability distribution functions for the vector
bosons \cite{Johnson:1987tj}. This improved formulation does not invoke
any kinematical approximation such as those mentioned in the previous
paragraph. The only approximation concerns the on shell continuation of
the vector boson scattering cross section. Using this
factorization technique and the relation between the polarization vectors
and the vector boson propagator in Unitary gauge, the matrix element
for any process of the kind $f_1f_2\rightarrow f_3f_4+Y$,
where $Y$ is produced by vector boson fusion, can be written as:

\be\label{M_tot}
M =e^2\sum_{m,n=-1}^{1} (-1)^{m+n}\frac{j_1(p_1,p_3).\epsilon^*_1(m)}
{q_1^2 - M_{V_1}^2}\times \frac{j_2(p_2,p_4).\epsilon^*_2(n)}
{q_2^2 - M_{V_2}^2}
\times M(m,n)
\ee

$q_{1,2}$ are the momenta of the initial vector bosons,
$\epsilon_j(m)$ are their
polarization vectors corresponding to the different helicity states
$m=0 , \pm 1$. We have used the same expressions, conventions and frame to
define them as given in \cite{Kuss:1995yv}. They are normalized according
to :

\be
\epsilon_j(m)\cdot\epsilon_j^*(m^\prime)=\delta_{m,m^\prime}(-1)^m
\ee
and satisfy the completeness relation:

\be
\sum_{m=-1,0,1} \epsilon_j^\mu(m) \epsilon_j^{*\nu}(m)= -g^{\mu\nu} +
\frac{q^\mu q^\nu}{M_{V_j}^2}
\ee

$j_{1,2}$ are the quark currents and $M(m,n)$ the off shell scattering
amplitude of the vector boson subprocess $V_1V_2\rightarrow V_3V_4$:

\be\label{M_ww_sca}
M(m,n)= \sum_{\mu \nu \alpha \beta} 
\epsilon_1^\mu(m)\epsilon_2^\nu(n)
T_{\mu \nu \alpha \beta} \epsilon_3^{*\alpha}  \epsilon_4^{*\beta}
\ee

Usually an integration is performed over all the integration variables
which are not concerned with the vector boson scattering subprocess as a
first step to compute the vector boson luminosities. Denoting these variables
by $\{\phi\}$, one can write the total cross section expression as:

\be
\sigma_{tot}= \int g(q_1^2, q_2^2, x, \phi) d\phi \:
\sigma^{VV}_{off}(W^2,q_1^2, q_2^2)dq_1^2 dq_2^2 dx
\ee

$ g(q_1^2, q_2^2, x, \phi)$ represents all terms which are independent of
the vector boson scattering subprocess. Here $x =\frac{W^2}{s_{qq}}$, where
$W^2 = (q_1 + q_2)^2$ is the diboson invariant mass squared. 

\be
\sigma^{VV}_{off}(W^2,q_1^2, q_2^2) \sim \int \sum_{m,n,m^\prime,n^\prime} 
M(m,n)M^*(m^\prime,n^\prime)dp_V
\ee  
is the off shell $VV\rightarrow VV$ cross section and $dp_V$ the final state
vector boson phase space element.
The next step is the extrapolation to on shell masses. In \cite{Kuss:1995yv}
it is achieved by simple proportionality factors between the off shell and
on shell cross sections:
\be
\sigma^{off}_{pol}(W^2, q_1^2, q_2^2)=f_{pol}(W^2, q_1^2, q_2^2)
\sigma^{on}_{pol}(W^2, M_{V_{1}}^2, M_{V_{2}}^2)  
\ee
the subscript $pol$ refers to the different vector boson polarization states,
and  $M_{V_{1}}$, $M_{V_{2}}$ the masses of the vector bosons initiating the
scattering process. While we refer to \cite{Kuss:1995yv} for the details,
we reproduce here the form factors 
$f_{pol}$ expression according to the different polarization configurations:

\bea\label{pol_fact}
f_{TT}=1 \: ; \:f_{LL}=(\frac{M_{V_{1}}^2}{-q_1^2})
(\frac{M^2_{V_{2}}}{-q_2^2})\\
f_{LT}=(\frac{M_{V_1}^2}{-q_1^2})\: ; \:f_{TL}=(\frac{M_{V_2}^2}{-q_2^2})\\
f_{TLTL}=(\frac{M_{V_1}}{\sqrt{-q_1^2}})(\frac{M_{V_2}}{\sqrt{-q_2^2}})
\eea
Finally one can write the cross section expression in the
EVBA approximation as: 
\be
\sigma^{EVBA}_{tot}= \int g(q_1^2, q_2^2, x, \phi) d\phi dq_1^2 dq^2_2
\sum_{Pol_{1}Pol_{2}}f_{pol} \sigma^{on}_{pol}dx
\ee
By comparison with eq.(\ref{evba_cros}) the luminosity can be expressed as:
\be
{\mathcal{ L}^{WW}_{Pol_{1}Pol_{2}}(x)}=\int g(q_1^2, q_2^2, x, \phi)
f_{pol} d\phi dq_1^2 dq^2_2 
\ee
In our implementation we have used the same assumptions but we have
performed the onshell extrapolation at the matrix
element level. This allows to keep all the terms
$M(m,n)M^*(m^{\prime},n^{\prime})$ ($m\neq m^\prime , n\neq n^\prime$)
in the total amplitude square expression. The offshell
vector boson scattering matrix 
element $M(m,n)$ in (\ref{M_tot}) and (\ref{M_ww_sca}) is expressed in terms
of the
corresponding on shell matrix elements and the polarization factors 
(\ref{pol_fact}) as 
\be\label{M_extrp}
M(m,n) = \sqrt{f_{mn}} M^{on}(m,n) \: \: \: \: \:(m,n=L,T).
\ee

Moreover we have not employed any luminosity function but we have
used the diagrammatical expression of the fermion lines, 
evaluating them for each kinematical configuration and polarization.

\begin{figure}[tbh]
\begin{center}
%\mbox{\epsfig{file=evba_exact_14tev.pdf,width=8cm}
%\epsfig{file=evba_exa_14_mh250.pdf,width=8cm}}
\mbox{\epsfig{file=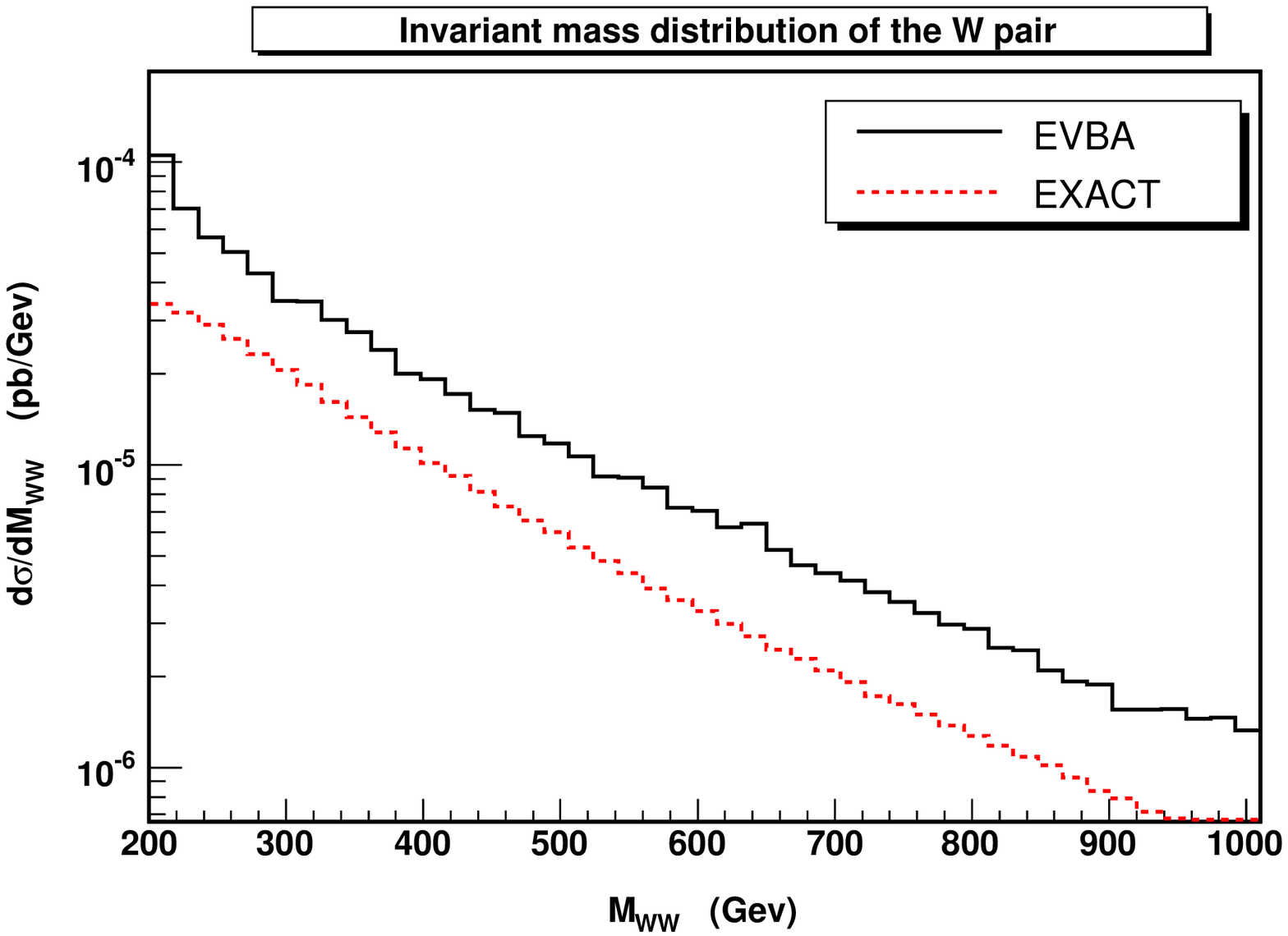,width=8cm}
\epsfig{file=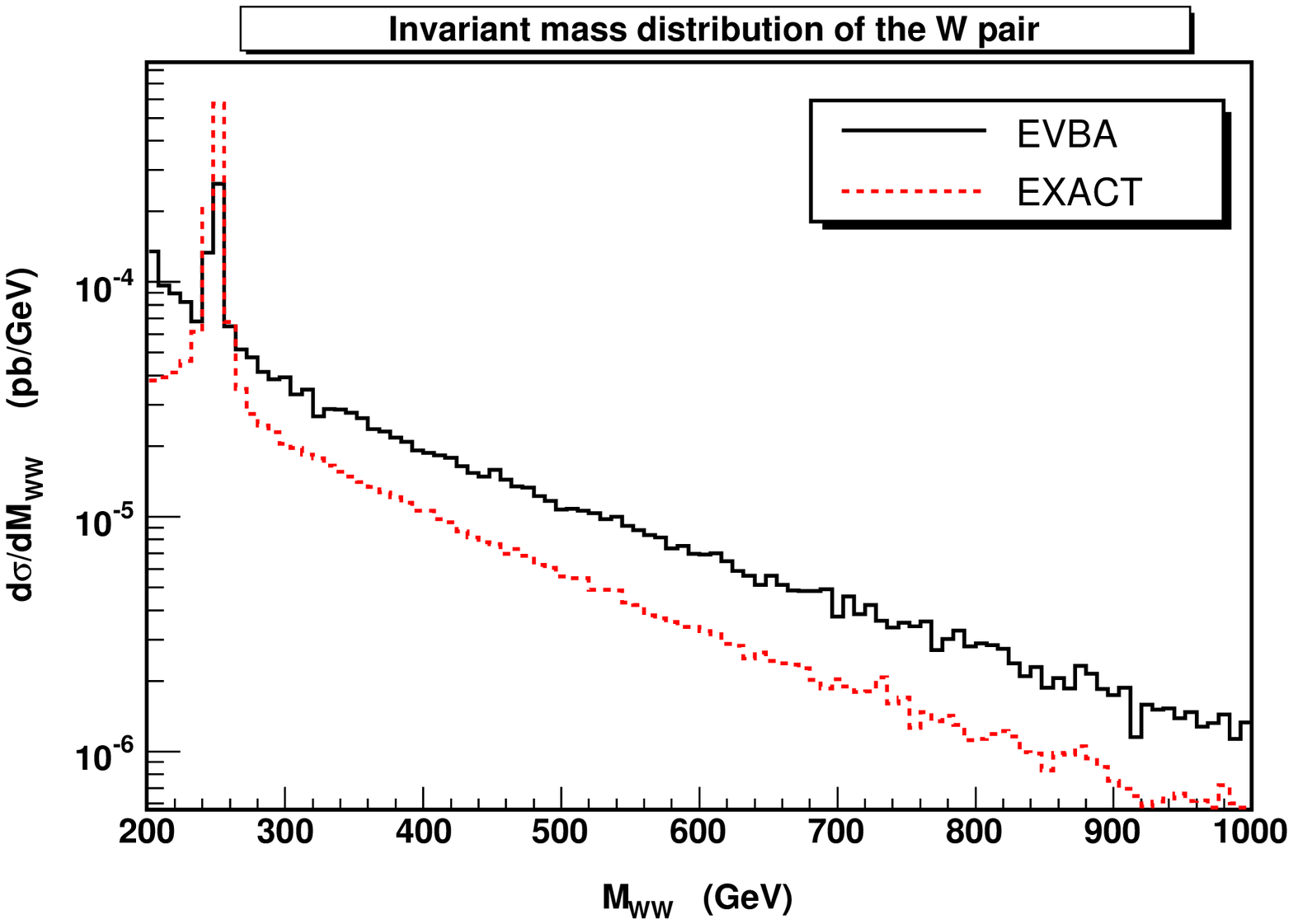,width=8cm}}
\caption{ $WW$ invariant mass distribution $M(WW)$ for the
process $us \rightarrow dc W^+W^-$  with EVBA
(black solid curve) and with exact complete computation (red dashed curve) for
the Infinite Higgs mass case (left) and $M_h$=250 GeV (right) at the LHC.
}
\label{Mww_evba_exa_LHC}
\end{center}
\end{figure}

\section{Comparison between EVBA and exact results}
\label{sec:EVBA-exa}
We have computed the total cross section and the distribution
of the vector boson pair invariant mass $M(WW)$ with an exact
calculation
and in EVBA for the process $PP\rightarrow us\rightarrow cd W^+W^-$ at
the LHC.
Cuts are necessary in EVBA to avoid the
photon $t$ channel propagator pole and the form factor (\ref{pol_fact})
($q_1^2\rightarrow 0 , q_2^2\rightarrow 0$) singularities.
We have restricted the CM
scattering angle of the W's and the laboratory frame
polar angle of the {\it c} and
{\it d} quarks in order to deal with photon and form factor singularities
respectively.
For comparison purpose, the same cuts have also been applied in the exact
calculation even though no singularity appears in this case.
Unless otherwise noted, we have used a cut of 10 degrees for all
three angles.

For the total cross section in the noHiggs case
we obtain 0.63 10$^{-2}$ $pb$ with an exact calculation 
and 1.36 10$^{-2}$ $pb$ in EVBA.
For $M_h$=250 GeV we obtain 1.32 10$^{-2}$ $pb$ and 1.64 10$^{-2}$ $pb$ 
respectively. The corresponding distributions of the $WW$ invariant mass is
shown fig. \ref{Mww_evba_exa_LHC}.
The ratio of the two results is .
The EVBA overestimates the exact calculation by a factor of about two, which is
almost insensitive to $M(WW)$,
with the exception of the Higgs peak region at $M_h$= 250 GeV where the exact
result is larger than the EVBA one.

We have checked our implementation of the EVBA against the one used in
{\tt PYTHIA} for the scattering of longitudinally polarized $W$'s. The two
results are only in qualitative agreement as expected both for the total cross
section and for distributions.

\begin{table}[htb]
\begin{center}
\begin{tabular}{|c|c|c|c|} 
\hline
%& \multicolumn{1}{|c|}{No-Higgs} \\
 $M_h$    &   EVBA $(pb)$ &   EXACT $(pb)$  & Ratio  \\
\hline
$\infty$ & 3.90 10$^{-2}$  & 1.78 10$^{-2}$   & 2.17 \\
\hline
130 GeV   & 3.94 10$^{-2}$  & 1.71 10$^{-2}$   & 2.3  \\
\hline
250 GeV   & 4.61 10$^{-2}$  & 4.09 10$^{-2}$   & 1.12 \\
\hline
500 GeV   & 4.42 10$^{-2}$  & 2.5 10$^{-2}$    & 1.77 \\
\hline
\end{tabular}
\caption{\footnotesize Total cross sections computed with EVBA and exact
computation and their ratio for the process  $us\rightarrow cd W^+W^-$ at fixed CM
energy $\sqrt{s} = 1$ TeV.}
\label{exa_evb_noh_1}
\end{center}
\end{table}

\begin{table}[tbh]
\begin{center}
\begin{tabular}{|c|c|c|c|} 
\hline
%& \multicolumn{1}{|c|}{No-Higgs} \\
 $\theta_{cut}$    &   EVBA $(pb)$ &   EXACT $(pb)$  & Ratio  \\
\hline
$10^\circ$   & 4.42 10$^{-2}$  &  2.5 10$^{-2}$    & 1.77 \\
$30^\circ$   & 1.33 10$^{-2}$  & 2.06 10$^{-2}$   & 0.64 \\
$60^\circ$   & 6.06 10$^{-3}$  & 1.28 10$^{-2}$   & 0.47 \\
\hline
\end{tabular}
\caption{\footnotesize Total cross section in EVBA and exact
computation and their ratio for different angular cuts.
The CM energy is $\sqrt{s} = 1$ TeV and the Higgs mass $M_h$=500 GeV.}
\label{exa_evb_mh500_1_theta}
\end{center}
\end{table}

In previous works \cite{Gunion:1986gm} the same partonic process
$us\rightarrow cd W^+W^-$  has been used for EVBA $vs$ exact comparisons
at fixed energy.
For this reason we have tried to reproduce the results of
\cite{Gunion:1986gm}, using as closely as possible the same cuts and parameters.
On one side this represented a test of our implementation of EVBA
and on  the
other side it allowed us to determine whether the EVBA results are improved
by this new and more rigorous formulation. We have computed the total cross
section
at a fixed center of mass energy of 1 TeV in the limit of infinite Higgs mass
and with Higgs masses $M_h$=130, 250, 500 GeV. 
In Tab. \ref{exa_evb_noh_1}  the cross sections and corresponding
ratio are reported.
We have also compared a
number of distributions obtained with the two versions of EVBA
and with the exact calculation. Our more sophisticated implementation doe not appear
to improve substantially the agreement with the exact results.

\begin{figure}[tbh]
\begin{center}
%\mbox{\epsfig{file=mh500_evba_exact_cuts_v3.pdf,width=9cm}}
\mbox{\epsfig{file=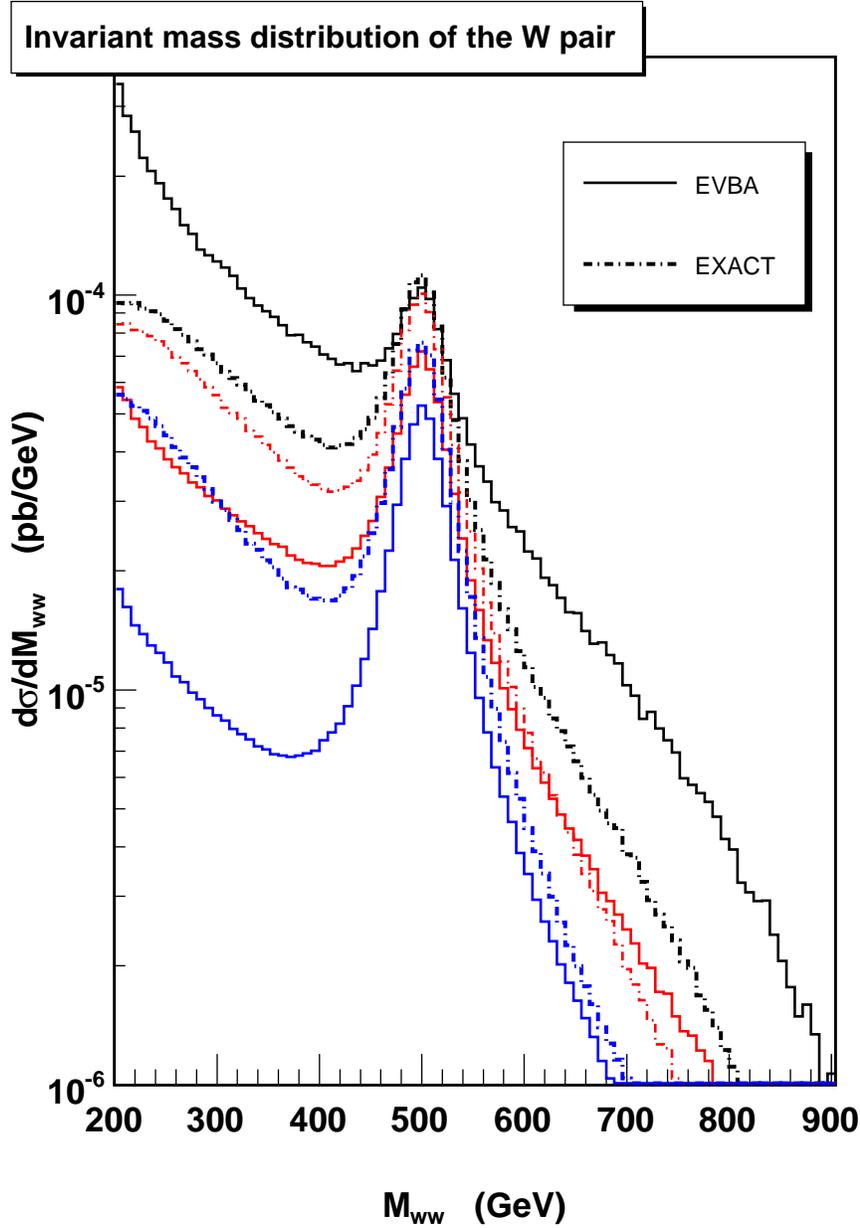,width=12cm}}
\caption{The $WW$ invariant mass distribution with different $\theta$-cut
with EVBA (solid curves) and with the exact calculation
(dashed curves). From top to bottom 
$\theta_c = 10^\circ$ (black), $30^\circ$ (red), $60^\circ$ (blue).
The CM energy is $\sqrt{s}=1$ TeV and $M_h$=500 GeV}
\label{Mww_evba_exa_1tev_cut30_60}
\end{center}
\end{figure}

Finally, we have analyzed the sensitivity of the results to the angular cut
in the $M_h$=500 GeV Higgs case.
The total cross section for $\theta_{cut}= \ 10^\circ,\ 30^\circ,\ 60^\circ$
are presented in Tab. \ref{exa_evb_mh500_1_theta}
which shows that the EVBA is more sensitive to the angular cut
than the exact computation.
The corresponding $M(WW)$ distribution is shown in 
fig \ref{Mww_evba_exa_1tev_cut30_60}.
We see that the relationship between the exact and EVBA results depends quite
appreciably on the angular cut.
The EVBA overestimates the exact result at $\theta_{cut}= \ 10^\circ$
by about a factor of two outside the Higgs peak, while the two distributions are
in fair agreement at the resonance.
However the EVBA underestimates the correct result
at $\theta_{cut}= \ 60^\circ$ over the whole mass range. The difference
decreases
from about a factor of two at small invariant masses to roughly 20\% at masses
larger than the Higgs mass.
Therefore, while it appears quite possibile to find a set of cuts, at fixed
energy and Higgs mass, for which the EVBA approximation reproduces well the
exact result for the total cross section,
in general it is extremely difficult to extract from the EVBA
more than a very rough estimate of the actual behaviour of the Standard Model
predictions for boson boson scattering.

\begin{figure}[tbh]
\begin{center}
%\mbox{\epsfig{file=noh_hig_cut_all_new.pdf,width=12cm}}
\mbox{\epsfig{file=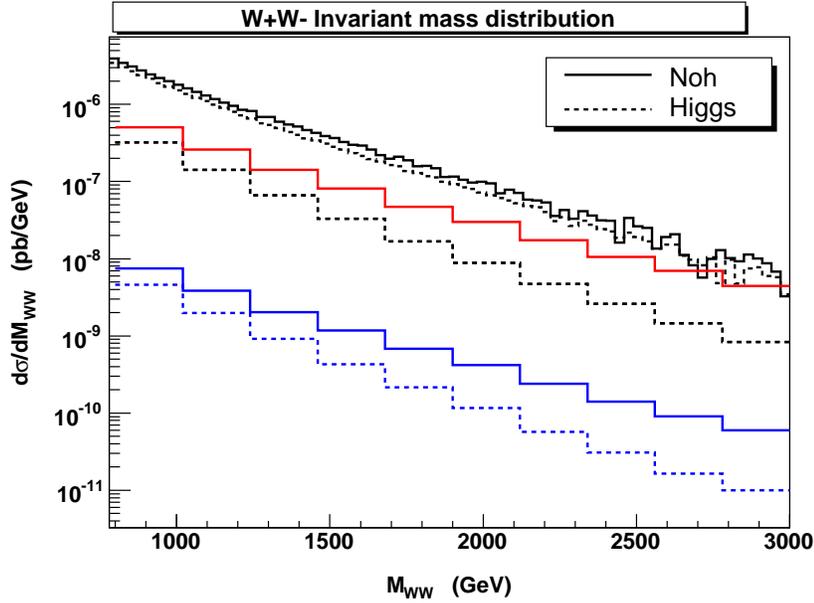,width=12cm}}
\caption{The $WW$ invariant mass distribution in
$PP\rightarrow us \rightarrow cdW^+W^-$ at the LHC
for the infinite Higgs mass case (solid curves) and for $ M_h$=200 GeV
(dashed curve). The two intermediate (red)
curves have been obtained imposing the set of cuts described in the text. 
The two lowest (blue) curves refer to the process 
$PP\rightarrow us \rightarrow cd\mu^-\overline{\nu}_\mu e^+\nu_e$; in this case
further acceptance cuts have been imposed on the charged leptons: $E_l > 20$
GeV,  $p_{T_l} > 10$ GeV , $\vert \eta_l \vert <$ 3.}
\label{noh_hig}
\end{center}
\end{figure}

\section{The large invariant mass region}
\label{sec:LargeMass}
The results presented in \sects{sec:GaugeInv}{sec:EVBA-exa}
lead to the conclusions that the
boson boson subamplitude cannot reliably be extracted from the full $qq
\rightarrow qqVV$ amplitude. However the full amplitude is sensitive to the
details of the EWSB mechanism. If no light Higgs is present in the SM spectrum,
some hitherto unknown mechanism must intervene to enforce unitarity of the
S-Matrix which embodies the conservation of total probability. The infinite
Higgs mass limit violates perturbative unitarity in onshell boson boson
scattering but on the other hand can be computed exactly, while the many
available models for unitarizing the theory deal exclusively with on shell
bosons and can only approximately be incorporated in a decription of
$qq \rightarrow qqVV$ processes or in a six final state fermion framework,
which has recently become available for the LHC
\cite{Accomando:2005cc,ballestrero:2006pn}.

\begin{table}[thb]
\begin{center}
\begin{tabular}{|c|}
\hline
      E(quarks)$>20$ GeV \\
\hline
      \pt(quarks,W)$>10$ GeV \\
\hline
      $2<\vert\eta$(quark)$\vert<6.5$ \\
\hline
      $\vert\eta$(W)$\vert<3$ \\
\hline
\end{tabular}
\caption{Selection cuts applied in \fig{noh_hig}.} 
\label{standard-cuts}
\end{center}
\end{table}
 
At the LHC the linear rise of the cross section at large boson boson invariant
masses squared entailed by the leading behaviour of boson boson scattering in
the SM with a very large Higgs mass will be overcome by the decrease of the parton
luminosities at large $x$ and will be particularly challenging to
detect. In the absence of a more reliable theory
we have adopted the noHiggs model as a poor man substitute.

In \fig{noh_hig} we show the large mass tail of the boson pair invariant
mass distribution for the noHiggs case and for a Higgs mass of 200 GeV.
In the absence of cuts the two results differ by about 20\% over the full range.
With appropriate cuts the difference between the two cases can be significantly
increased. Applying the selection cuts in \tbn{standard-cuts} we obtain the two
intermediate, red curves in \fig{noh_hig}.
The two lowest (blue) curves refer to the full process 
$PP\rightarrow us \rightarrow cd\mu^-\overline{\nu}_\mu e^+\nu_e$; in this case
further acceptance cuts have been imposed on the charged leptons: $E_l > 20$
GeV,  $p_{T_l} > 10$ GeV , $\vert \eta_l \vert <$ 3.
We see that the separation between the two Higgs mass hypotheses persists also
in this more realistic setting.

\begin{figure}[tbh]
\begin{center}
%\mbox{\epsfig{file=delta_eta_noh_hig_v2.pdf,width=12cm}}
\mbox{\epsfig{file=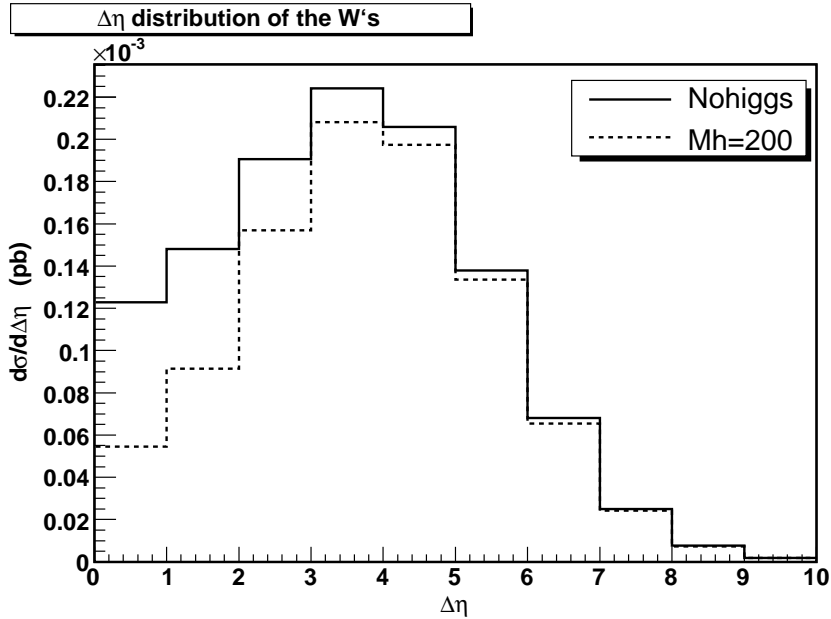,width=12cm}}
\caption{ Distribution of the pseudorapidity difference between the
two W's for 
$us \rightarrow dc W^+W^-$ at the LHC  
for the Infinite Higgs mass case (solid curve) and $M_h$=200 GeV
(dashed curve).
}
\label{Deltaeta_noh_hig}
\end{center}
\end{figure}

As an example of the different kinematical distributions in the two cases we
show
in \fig{Deltaeta_noh_hig} the absolute value of the difference between the
pseudorapidities of the two $W$'s.
It is interesting that the shape of the kinematical distributions,
which are less sensitive to pdf uncertainties
than their absolute normalization,
behave differently when a light Higgs is present in the spectrum then when
its mass is very large.
For more details we refer to \cite{Accomando:2005hz,Accomando:2006vj}.
We conclude that while it does not appear to be possible to study the
contribution of the scattering diagrams in isolation from the remaining ones,
the full amplitude, in the region of
large $WW$ invariant masses at LHC energies, is sensitive to the presence of
a light Higgs and therefore to the details of the mechanism of EWSB.

\section{Conclusions}
We have critically examined the role of gauge invariance
in \VV -fusion processes and the reliability of the EVBA in describing them
in the Unitary, Feynman and Axial gauge.
We have shown that the $WW$ scattering diagrams
do not constitute the dominant contribution in any phase space region for the
set of gauge fixing we have examined.
The Axial gauge as proposed in \cite{Kunszt:1987tk} results in
less severe cancellations between the contribution of non scattering diagrams 
and the contribution of the $WW$ scattering ones but typically the two sets
have comparable magnitude.

We have shown that EVBA results and their relationship to exact results
depend quite sensitevely on the set of cuts
which need to be applied in order to obtain a finite result.
Therefore, it is extremely difficult to extract from the EVBA
more than a very rough estimate of the actual behaviour of the Standard Model
predictions for boson boson scattering.

We conclude that while it seems impossible to isolate the
contribution of the scattering diagrams, the mechanism of EWSB
can be investigated, using the full amplitude, by a careful analysis
of the region of large $WW$ invariant masses at the LHC.

\bibliography{bibliography}

\newpage

\appendix
\section{Axial gauge}
\label{app:A}
{
We report here for convenience the main formulae describing the Axial gauge
formulation of \cite{Kunszt:1987tk} to which we refer for a more detailed
discussion.

Parametrizing the Higgs field as
\begin{displaymath}
 \phi (x) = \left( \begin{array}{c}
   iw(x) \\
   \sqrt{\frac{1}{2}} [ v + h(x) +iz(x) ]
   \end{array}  \right)
\end{displaymath}
the propagators for the five dimensional vector field ($W^\mu,w$) is
\be
\label{propW}
i\Delta^{IJ} = \frac{i}{q^2 - M_W^2}\it{N}^{IJ}(q)
\ee
where
\be
{N}^{\mu\nu}(q) = - g^{\mu\nu} + \frac {q^\mu n^\nu + n^\mu q^\nu}{q\cdot n}
                 - \frac {n^2}{(q\cdot n)^2}q^\mu q^\nu
\ee
\be
{N}^{\mu s}(q) = -i M_W  \frac {q\cdot n n^\nu + n^2 q^\mu}{(q\cdot n)^2}
\ee
\be
{N}^{s \nu }(q) = {N}^{ \nu s}(q)^\ast ={N}^{ \nu s}(-q)
\ee
\be
{N}^{s s }(q) = 1 - \frac {M_W^2 n^2}{(q\cdot n)^2}
\ee
and the index $s$ indicates the scalar component.

The polarization vectors satisfy
\be
\it{N}^{IJ}(q) = \sum_{\lambda = 1,2,L}
   \epsilon^I(q,\lambda ) \epsilon^J(q,\lambda )^\ast
\ee
For $\lambda = 1,2$  they describe transverse polarization:
\be
\epsilon^s(q,\lambda ) = 0 \qquad q_\mu \epsilon^\mu(q,\lambda ) =0 \qquad
n_\mu \epsilon^\mu(q,\lambda ) =0 \qquad
\epsilon_\mu(q,\lambda )\epsilon^\mu(q,\lambda^\prime )^\ast = 
- \delta_{\lambda \lambda^\prime }
\ee
while for $\lambda = L$  they describe longitudinal polarization:
\be
\epsilon^s(q,L ) = -i \sqrt{1 - M_W^2 n^2/(q\cdot n)^2} 
\ee
\be
\label{pol_s_mu}
\epsilon^\mu(q,L ) = \frac{- [M_W/q\cdot n] n^\mu + [M_W^2 n^2/(q\cdot
n)^2]q^\mu}{\sqrt{1 - M_W^2 n^2/(q\cdot n)^2}}
\ee
The propagators and polarizations for the ($Z^\mu,w$) field
can be obtained from \eqns{propW}{pol_s_mu} with the substitution
$M_W \rightarrow M_Z$.

The remaining Feynman rules are identical to the ones in R$_\xi$ gauges.
In our calculation we have used $\xi = 1$.
} %end appendix

\end{document}